\documentclass[12pt]{article}

\begin{document}

\thispagestyle{empty}
\renewcommand{\thefootnote}{\fnsymbol{footnote}}

\begin{flushright}
{\small
SLAC--PUB--8676\\
October, 2000\\}
\end{flushright}

\vspace{.8cm}

\begin{center}
{\bf\Large   
Nonlinear $\delta f$ Method for Beam-Beam Simulation 
\footnote{Work supported by
Department of Energy contract  DE--AC03--76SF00515.}}

\vspace{1cm}

Yunhai Cai, Alexander W. Chao, Stephan I. Tzenov\\
Stanford Linear Accelerator Center, Stanford University,
Stanford, CA  94309\\

\medskip

and\\

\medskip

Toshi Tajima\\
University of Texas at Austin, Austin, TX 78712\\
and\\
Lawrence Livermore National Laboratory, Livermore, CA 94551\\
\end{center}

\vfill

\begin{center}
{\bf\large  
Abstract }
\end{center}

\begin{quote}
We have developed an efficacious algorithm for simulation 
of the beam-beam interaction in synchrotron colliders based on 
the nonlinear $\delta f$ method, where $\delta f$ is the much 
smaller deviation of the beam distribution from the slowly 
evolving main distribution $f_0$. In the presence of damping 
and quantum fluctuations of synchrotron radiation it has been 
shown that the slowly evolving part of the distribution function 
satisfies a Fokker-Planck equation. Its solution has been 
obtained in terms of a beam envelope function and an amplitude 
of the distribution, which satisfy a coupled system of ordinary 
differential equations. A numerical algorithm suited for direct 
code implementation of the evolving distributions for 
both $\delta f$ and $f_0$ has been developed. Explicit 
expressions for the dynamical weights of macro-particles for 
$\delta f$ as well as an expression for the slowly changing
$f_0$  have been obtained.
\end{quote}

\vfill

\begin{center} 
{\it Submitted to} {\bf Physical Review Special Topics: 
Accelerators and Beams} 
\end{center}

\newpage


\pagestyle{plain}

\renewcommand{\theequation}{\thesection.\arabic{equation}}

\setcounter{equation}{0}

\section{Introduction}

The effects of the beam-beam interaction on particle dynamics in a 
synchrotron collider are the key element that determines the 
performance of the collider such as luminosity \cite{chao} - 
\cite{month}. In order to accurately understand these effects, it 
is necessary to incorporate not only the overall collisional 
effects of the beam-beam interaction, but also the collective 
interaction among individual parts of the beam in each beam and its 
feedback on the beam distribution. The particle-in-cell (PIC) approach 
\cite{birdsall}, \cite{tajima} has been adopted to address such a 
study need \cite{krish1}, \cite{krish2}, \cite{cai}. 

Particle-in-cell codes typically use macro-particles to 
represent the entire distribution of particles. In the beam-beam 
interaction for the PEP-II \cite{book} (for example), the beams 
consist of $10^{10}$ particles each. Simulating this many particles 
with the PIC technique is computationally prohibitive. With the 
conventional PIC code $10^{10}$ particles are represented by 
only $10^3 - 10^4$ macro-particles allowing simulation of 
the beam-beam interaction in a reasonable computation time. 
However, the statistical fluctuation level of various quantities such as 
the beam density $\rho$ in the code is much higher than that of 
the real beam. The fluctuation level $\delta \rho$ goes as 
approximately

\begin{eqnarray}
{\frac {\delta \rho} {\rho}} \approx 
{\frac {\sqrt{N}} {N}}, 
\label{Fluclev} 
\end{eqnarray}

\noindent
where $N$ is the number of particles. Therefore, the fluctuation 
level of the PIC code is about $10^3$ times higher than that of 
the real beam. Although this probability is not significant for 
beam blowup near resonances, the higher fluctuation level has a 
large effect on more subtle phenomenon such as particle 
diffusion. The purpose of the $\delta f$ algorithm is to facilitate 
the study of subtle effects and has been introduced in \cite{perkins}, 
\cite{kots}, \cite{koga}.

The $\delta f$ method follows only the fluctuating part of the 
distribution instead of the entire distribution. This is 
essentially modeling the numerator of the right-hand side of 
equation (\ref{Fluclev}). So the $10^3 - 10^4$ macro 
particles are used to represent ${\sqrt{10^{10}}}$ or $10^5$ 
real fluctuation particles in PEP-II beams. This is only one 
or two orders of magnitude beyond the number of macro 
particles. Such a modest gap between the number of macro
particles and the real fluctuating particles maybe ameliorated 
by the standard techniques of the PIC approach, such as the 
method of finite-sized macro-particles \cite{birdsall}, 
\cite{tajima}.

PIC strong-strong codes use a finite number of particles to 
represent the Klimontovich equation for the microscopic phase 
space density (MPSD) \cite{klimon}. In the particular case of 
one-dimensional beam-beam interaction, 

\begin{eqnarray}
{\frac {\partial f} {\partial s}} + 
p {\frac {\partial f} {\partial x}} - 
{\left( K(s) x - F(x; s) \right)} 
{\frac {\partial f} {\partial p}} = 0, 
\label{Klimont} 
\end{eqnarray}

\noindent
where $K(s) x$ is the usual magnetic guiding force and $F(x; s)$ 
is the beam-beam force

\begin{eqnarray}
F(x; s) = {\frac {2 e E_x (x)} {m \gamma v^2}} 
\delta_p (s). 
\label{Bbfor} 
\end{eqnarray}

\noindent
The electric field $E_x (x)$ is calculated from the distribution 
of the particles of the on-coming beam and $\delta_p (s)$ is the periodic 
$\delta$-function with a periodicity of the accelerator circumference. 
The distribution function $f(x, p; s)$ is represented by a finite 
number of macro-particles by 

\begin{eqnarray}
f(x, p; s) = {1\over N} \sum \limits_{n=1}^N  
\delta (x - x_n (s)) \delta (p - p_n (s)), 
\label{Distr} 
\end{eqnarray}

\noindent
where $N$ is the number of macro-particles. 

The strategy of the $\delta f$ method is that only the 
perturbative part of the distribution is followed. The total 
distribution function $f(x, p; s)$ is decomposed into

\begin{eqnarray}
f(x, p; s) = f_0(x, p; s) + \delta f(x, p; s), 
\label{Decomp} 
\end{eqnarray}

\noindent
where $f_0(x, p; s)$ is the steady or slowly varying part of the 
distribution and $\delta f(x, p; s)$ is the perturbative part. The 
key to this method is finding a distribution $f_0(x, p; s)$ which 
is close to the total distribution $f(x, p; s)$. The perturbative 
part $\delta f(x, p; s)$ is then small, causes only small changes 
to the distribution, and thus represents only the fluctuation 
levels. If a distribution $f_0(x, p; s)$ close to the total 
distribution is not found or found poorly, then $\delta f(x, p; s)$ 
represents more than the fluctuation part of the total distribution; 
defeating the purpose of the method. The ideal situation is having 
an analytic solution for $f_0(x, p; s)$. In this case any numerical 
truncation errors which result from the necessary derivatives of 
this function are eliminated. If an analytic solution cannot be 
found, then a numerical solution needs to be found which is close 
to the total distribution $f(x, p; s)$ and is slowly varying. A 
frequent numerical update of $f_0(x, p; s)$ would also defeat the 
purpose of the $\delta f$ method, since the PIC technique 
essentially does this also.

The beam-beam interaction can lead to beam instabilities that 
disrupt or severely distort the beam or gradual beam spreading. The 
higher the beam current, and thus the beam-beam interaction, the 
stronger these effects become. Therefore, when one wants to 
maximize the luminosity of a collider, one needs to confront the 
beam-beam interaction effects. The operation of PEP-II, for example, 
is critically dependent on the beam-beam interaction and optimal 
parameters to minimize the related beam instabilities are under intense
study. 

The paper is organized as follows. In the next Section we present a 
brief formulation of the problem of beam-beam interaction in 
synchrotron colliders. In Section 3 we develop the nonlinear $\delta f$ 
method for solving the equation for the microscopic phase space density 
in the presence of random external forces. The equation for the 
fluctuating part $\delta f$ is being derived and its solution is found 
explicitly in terms of dynamical weight functions, prescribed to each 
macro-particle. In Section 4 we solve the Fokker-Planck equation for the 
averaged slowly evolving part of the distribution. We show that the 
solution is an exponential of a bilinear form in coordinates and 
momenta with coefficients that can be regarded as generalized 
Courant-Snyder parameters. In Section 5 we outline numerical 
algorithms to alternatively solve the Fokker-Planck equation and
the macro particle distribution with dynamical weight. Finally, 
Section 6 is dedicated to our summary and conclusions.

\setcounter{equation}{0}

\section{Description of the beam-Beam Interaction}

In order to describe the beam dynamics in an electron positron 
storage ring, we introduce the equations of motion in the following 
manner. The beam propagation in a reference frame attached to the 
particle orbit is usually described in terms of the canonical 
conjugate pairs

\begin{eqnarray}
{\widehat{u}}^{(k)} = u^{(k)} - D_u^{(k)}{\widehat{\eta}^{(k)}} 
\qquad ; \qquad 
{\widehat{p}}_u^{(k)} = {\frac {p_u^{(k)}} {p_0^{(k)}}} 
- {\widehat{\eta}^{(k)}} {\frac {dD_u^{(k)}} {ds}}, 
\label{Conj1} 
\end{eqnarray}

\begin{eqnarray}
{\widehat{\sigma}}^{(k)} = {\widetilde{\sigma}}^{(k)} + 
{\sum\limits_{u=x,z}} {\left( u^{(k)} {\frac {dD_u^{(k)}} {ds}} 
- D_u^{(k)} {\frac {p_u^{(k)}} {p_0^{(k)}}} \right)} 
\qquad ; \qquad 
{\widehat{\eta}}^{(k)} = {\frac {1} {\beta_{k0}^2}} 
{\frac {E^{(k)} - E_{k0}} {E_{k0}}}, 
\label{Conj2} 
\end{eqnarray}

\noindent
where $u = (x,z)$, $s$ is the path length along the particle orbit, 
and the index $k$ refers to either beam $(k = 1,2)$. In equations 
(\ref{Conj1}) and (\ref{Conj2}) the quantity $u^{(k)}$ is the 
actual particle displacement from the reference orbit in the plane 
transversal to the orbit, $p_u^{(k)}$ is the actual particle 
momentum, and $E^{(k)}$ is the particle energy. Furthermore, 
$p_0^{(k)}$ and $E_{k0}$ are the total momentum and energy of the 
synchronous particle, respectively, and $D_u^{(k)}$ is the 
well-known dispersion function. The quantity

\begin{eqnarray}
{\widetilde{\sigma}}^{(k)} = s - \omega_0^{(k)} R t 
\label{LongDispl} 
\end{eqnarray}

\noindent
is the longitudinal coordinate of a particle from the $k$-th beam 
with respect to the synchronous particle, where $\omega_0^{(k)}$ is 
the angular frequency of the synchronous particle and $R$ is the 
mean machine radius.

It is known that the dynamics of an individual particle is governed 
by the Langevin equations of motion:

\begin{eqnarray}
{\frac {d{\widehat{u}}^{(k)}} {ds}} = 
{\frac {\partial {\widehat{H}}^{(k)}} 
{\partial {\widehat{p}}_u^{(k)}}} - D_u^{(k)} 
{\widetilde{F}}_{\eta}^{(k)} 
\qquad ; \qquad 
{\frac {d{\widehat{p}}_u^{(k)}} {ds}} = 
- {\frac {\partial {\widehat{H}}^{(k)}} 
{\partial {\widehat{u}}^{(k)}}} + {\widetilde{F}}_u^{(k)} 
- {\widetilde{F}}_{\eta}^{(k)} {\frac {dD_u^{(k)}} {ds}}, 
\label{Langevin1} 
\end{eqnarray}

\begin{eqnarray}
{\frac {d{\widehat{\sigma}}^{(k)}} {ds}} = 
{\frac {\partial {\widehat{H}}^{(k)}} 
{\partial {\widehat{\eta}}^{(k)}}} - {\sum\limits_{u=x,z}} 
D_u^{(k)} {\widetilde{F}}_{u}^{(k)} 
\qquad ; \qquad 
{\frac {d{\widehat{\eta}}^{(k)}} {ds}} = 
- {\frac {\partial {\widehat{H}}^{(k)}} 
{\partial {\widehat{\sigma}}^{(k)}}} + 
{\widetilde{F}}_{\eta}^{(k)}, 
\label{Langevin2} 
\end{eqnarray}

\noindent
where

\begin{eqnarray}
{\widetilde{F}}_u^{(k)} = - p_0^{(k)}{\cal A}_k 
{\left( {\widehat{p}}_u^{(k)} + {\widehat{\eta}}^{(k)} 
{\frac {dD_u^{(k)}} {ds}} \right)}, 
\label{Sforceu} 
\end{eqnarray}

\begin{eqnarray}
{\widetilde{F}}_{\eta}^{(k)} = - p_0^{(k)}{\cal A}_k 
{\left[ 1 + {\left( 3 - {\beta}_{k0}^2 + {\alpha}_M^{(k)} 
\right)} {\widehat{\eta}}^{(k)} + 
{\sum\limits_{u=x,z}} K_u^{(k)} 
{\widehat{u}}^{(k)} \right]}, 
\label{Sforceeta} 
\end{eqnarray}

\begin{eqnarray}
{\cal A}_k = {\cal C}_1 {\left| {\bf B}_k \right|}^2 + 
{\sqrt{{\cal C}_2}} {\left| {\bf B}_k \right|}^{3/2} 
\xi_k {\left( s \right)}, 
\label{Akcoeff} 
\end{eqnarray}

\begin{eqnarray}
{\cal C}_1 = {\frac {2 r_e e^2} {3 {\left( m_e c \right)}^3}} 
\qquad ; \qquad 
{\cal C}_2 = {\frac {55} {24 {\sqrt{3}}}} 
{\frac {r_e \hbar e^3} {{\left( m_e c \right)}^6}} 
\qquad ; \qquad 
r_e = {\frac {e^2} {4 \pi \epsilon_0 m_e c^2}}. 
\label{C12coeff} 
\end{eqnarray}

\noindent
Here $\alpha_M^{(k)}$ is the momentum compaction factor, 
$K_u^{(k)} {\left( s \right)}$ is the local curvature of the 
reference orbit, and ${\bf B}_k = {\left( B_x^{(k)}, 
B_z^{(k)}, B_s^{(k)} \right)}$ is the magnetic field. The variable 
${\xi_k {\left( s \right)}}$ is a Gaussian random variable with 
formal properties:

\begin{eqnarray}
{\left\langle \xi_k {\left( s \right)} 
\right\rangle} = 0 
\qquad ; \qquad 
{\left\langle \xi_k {\left( s \right)} 
\xi_k {\left( s' \right)} \right\rangle} = 
\delta {\left( s - s' \right)}. 
\label{Xi} 
\end{eqnarray}

\noindent
The hamiltonian part in equations (\ref{Langevin1}) and 
(\ref{Langevin2}) consists of three terms:

\begin{eqnarray}
{\widehat{H}}^{(k)} = {\widehat{H}}_0^{(k)} + 
{\widehat{H}}_2^{(k)} + {\widehat{H}}_{BB}^{(k)}, 
\label{Hamiltonian} 
\end{eqnarray}

\noindent
where

\begin{eqnarray}
{\widehat{H}}_0^{(k)} = - {\frac{{\cal K}^{(k)}} {2}} 
{\widehat{\eta}}^{(k)2} + 
{\frac{1} {2 \pi \beta_{k0}^2 }} 
{\frac{\Delta E_{k0}} {E_{k0}}} {\cos 
{\left( {\frac{h_k {\widetilde{\sigma}}_k } {R}} + 
\Phi_{k0} \right)}}, 
\label{Ham0} 
\end{eqnarray}

\begin{eqnarray}
{\widehat{H}}_2^{(k)} = {\frac{1} {2}} {\left( 
{\widehat{p}}_x^{(k)2} + {\widehat{p}}_z^{(k)2} \right)} + 
{\frac{1} {2R^2}} {\left( G_x^{(k)} {\widehat{x}}^{(k)2} + 
G_z^{(k)} {\widehat{z}}^{(k)2} \right)}, 
\label{Ham2} 
\end{eqnarray}

\begin{eqnarray}
{\widehat{H}}_{BB}^{(k)} = \lambda_k \delta_p 
{\left( s \right)} V_k {\left( x^{(k)}, z^{(k)}, 
{\widetilde{\sigma}}^{(k)}; s \right)}. 
\label{Hambb} 
\end{eqnarray}

\noindent
The parameter ${\cal K}^{(k)}$ is the so called slip phase 
coefficient, $h_k$ is the harmonic number of the RF field and 
$\Delta E_{k0}$ is the energy gain per turn. The coefficients 
$G_{x,z}^{(k)} {\left( s \right)}$ represent the focusing 
strength of the linear machine lattice, 
$\delta_p {\left( s \right)}$ is the periodic delta-function, 
while $\lambda_k$ and $V_k {\left( x^{(k)}, z^{(k)}, 
{\widetilde{\sigma}}^{(k)}; s \right)}$ are the beam-beam coupling 
coefficient and the beam-beam potential, respectively. The latter 
are given by the expressions:

\begin{eqnarray}
\lambda_k = {\frac{r_e N_{3-k}} {\gamma_{k0}}} 
{\frac{1 + \beta_{k0} \beta_{(3-k)0}} {\beta_{k0}^2}}, 
\label{Lambda} 
\end{eqnarray}

\begin{eqnarray}
V_k {\left( x^{(k)}, z^{(k)}, 
{\widetilde{\sigma}}^{(k)}; s \right)} = {\int} dx dz 
d{\widetilde{\sigma}} {\cal G}_k {\left( u^{(k)} - u, 
{\widetilde{\sigma}}^{(k)} - {\widetilde{\sigma}}; s
\right)} 
\rho_{3-k} {\left( u, {\widetilde{\sigma}}; s \right)}, 
\label{Potentialbb} 
\end{eqnarray}

\noindent
where $N_k$ is the number of particles in the $k$-th beam and 
the Green's function ${\cal G}_k {\left( u, {\widetilde{\sigma}}; 
s \right)}$ for the Poisson equation in the fully 3D case, in the 
ultra-relativistic 2D case and in the 1D case can be written respectively 
as:

\begin{eqnarray}
{\cal G}_k {\left( u^{(k)} - u, 
{\widetilde{\sigma}}^{(k)} - {\widetilde{\sigma}}; 
s \right)} = \left\{ \begin{array}{ll} 
      - {\left[ {\left( x^{(k)} - x \right)}^2 + 
      {\left( z^{(k)} - z \right)}^2 + 
      {\left( {\widetilde{\sigma}}^{(k)} - 
      {\widetilde{\sigma}} +2 s \right)}^2 \right]}
      ^{-1/2}, \\ 
               \\ 
   \delta {\left( {\widetilde{\sigma}}^{(k)} - 
   {\widetilde{\sigma}} +2 s \right)} 
   \ln {\left[ {\left( x^{(k)} - x \right)}^2 + 
   {\left( z^{(k)} - z \right)}^2  \right]}, \\ 
               \\ 
      2 \pi \delta {\left( {\widetilde{\sigma}}^{(k)} 
      - {\widetilde{\sigma}} +2 s \right)} 
      \delta {\left( z^{(k)} - z \right)} 
      {\left| x^{(k)} - x \right|}. 
\end{array} 
\right. 
\label{Green} 
\end{eqnarray}

In what follows we focus on the two-dimensional case, entirely 
neglecting the longitudinal dynamics. Let us write down the 
Langevin equations of motion (\ref{Langevin1}) and 
(\ref{Langevin2}) once again in the following form:

\begin{eqnarray}
{\frac {d{\bf x}^{(k)}} {ds}} = 
{\bf p}^{(k)}, 
\label{Lan1} 
\end{eqnarray} 

\begin{eqnarray}
{\frac {d{\bf p}^{(k)}} {ds}} = 
{\bf F}_L^{(k)} + {\bf F}_B^{(k)} + 
{\bf F}_R^{(k)}, 
\label{Lan2} 
\end{eqnarray}

\begin{eqnarray}
{\bf x}^{(k)} = {\left( {\widehat{x}}^{(k)} \; , 
\; {\widehat{z}}^{(k)} \right)} \qquad ; \qquad 
{\bf p}^{(k)} = {\left( {\widehat{p}}_x^{(k)} \; , 
\; {\widehat{p}}_z^{(k)} \right)}, 
\label{Vecvar} 
\end{eqnarray}

\noindent
where

\begin{eqnarray}
{\bf F}_L^{(k)} = {\left( - 
{\frac {G_x^{(k)}} {R^2}} {\widehat{x}}^{(k)} \; , 
\; - {\frac {G_z^{(k)}} {R^2}} {\widehat{z}}^{(k)} 
\right)} 
\label{Forcel} 
\end{eqnarray}

\noindent
is the (external) force acting on particles from the $k$-th 
beam, that is due to the linear focusing properties of the 
corresponding confining lattice. Furthermore, 

\begin{eqnarray}
{\bf F}_B^{(k)} = \lambda_k \delta_p (s) 
{\left( - {\frac {\partial V_k} 
{\partial {\widehat{x}}^{(k)}}}  \; , 
\; - {\frac {\partial V_k} 
{\partial {\widehat{z}}^{(k)}}} \right)} 
\label{Forceb} 
\end{eqnarray}

\noindent
is the beam-beam force and 

\begin{eqnarray}
{\bf F}_R^{(k)} = - p_{k0} {\cal A}_k 
{\left( {\widehat{p}}_x^{(k)} - 
{\frac {dD_x^{(k)}} {ds}} \; , \; 
{\widehat{p}}_z^{(k)} - 
{\frac {dD_z^{(k)}} {ds}} \right)} 
\label{Forcer} 
\end{eqnarray}

\noindent
is the synchrotron radiation friction force with a stochastic 
component due to the quantum fluctuations of synchrotron 
radiation [cf expression (\ref{Akcoeff})]. 

\setcounter{equation}{0}

\section{The Nonlinear $\delta f$ Method}

It can be checked in a straightforward manner that the 
Klimontovich microscopic phase space density 

\begin{eqnarray}
f_k {\left( {\bf x}, {\bf p}; s \right)} = 
{\frac {1} {N_k}}
\sum \limits_{n=1}^{N_k} \delta 
{\left[ {\bf x} - {\bf x}_n^{(k)} (s) \right]} 
\delta {\left[ {\bf p} - {\bf p}_n^{(k)} (s) 
\right]} 
\label{Mpsd} 
\end{eqnarray}

\noindent
satisfies the following evolution equation:

\begin{eqnarray}
{\frac {\partial f_k} {\partial s}} + {\bf p} 
\cdot {\bf \nabla}_x f_k + {\left( {\bf F}_L^{(k)} + 
{\bf F}_B^{(k)} \right)} \cdot {\bf \nabla}_p f_k + 
{\bf \nabla}_p \cdot {\left( {\bf F}_R^{(k)} f_k 
\right)} = 0, 
\label{Liouvk} 
\end{eqnarray}

\noindent
where ${\left\{ {\bf x}_n^{(k)} (s) \; , \; 
{\bf p}_n^{(k)} (s)  \right\}}$ is the trajectory of the $n$-th 
particle from the $k$-th beam. Next we split the MPSD $f_k$ 
into two parts according to the relation:

\begin{eqnarray}
f_k {\left( {\bf x}, {\bf p}; s \right)} = 
f_{k0} {\left( {\bf x}, {\bf p}; s \right)} + 
\delta f_k {\left( {\bf x}, {\bf p}; s \right)}, 
\label{Defdelf} 
\end{eqnarray}

\noindent
where $f_{k0}$ is a solution to the equation

\begin{eqnarray}
{\frac {\partial f_{k0}} {\partial s}} + {\bf p} 
\cdot {\bf \nabla}_x f_{k0} + {\left( {\bf F}_L^{(k)} + 
{\bf F}_{L0}^{(k)} \right)} \cdot {\bf \nabla}_p f_{k0} + 
{\bf \nabla}_p \cdot {\left( {\bf F}_R^{(k)} f_{k0} 
\right)} = 0. 
\label{Fequat0} 
\end{eqnarray}

\noindent
The quantity ${\bf F}_{L0}^{(k)}$ in Eq. 
(\ref{Fequat0}) is the {\it linear} part of the beam-beam force 
${\bf F}_B^{(k)}$. The beam-beam force should be calculated with the
on-coming beam distribution $f_{(3-k)0}$. In what follows it will 
prove convenient to cast the beam-beam force into the form:

\begin{eqnarray}
{\bf F}_B^{(k)} = {\bf F}_{L0}^{(k)} + 
{\bf F}_{N0}^{(k)} + \delta {\bf F}_B^{(k)}, 
\label{Bbforce} 
\end{eqnarray}

\noindent
where ${\bf F}_{N0}^{(k)}$ is the nonlinear (in the transverse 
coordinates) contribution calculated with $f_{(3-k)0}$, while 
$\delta {\bf F}_B^{(k)}$ denotes the part of the beam-beam 
force due to $\delta f_{3-k}$. 

It is worthwhile to note here that the representation 
(\ref{Defdelf}) is unique, embedding the basic idea of the 
$\delta f$ method. However, one is completely free to fix the 
$f_0$ part, which usually describes those features of the 
evolution of the system one can solve easily (and preferably 
in explicit form). In the next Section we show that $f_{k0}$, 
averaged over the statistical realizations of the process 
$\xi_k (s)$ satisfies a Fokker-Planck equation and find its 
solution.

Subtract now the two equations (\ref{Liouvk}) and 
(\ref{Fequat0}) to obtain an equation for the $\delta f_k$ 

\begin{eqnarray}
{\frac {\partial \delta f_k} {\partial s}} + {\bf p} 
\cdot {\bf \nabla}_x \delta f_k + {\left( {\bf F}_L^{(k)} + 
{\bf F}_B^{(k)} \right)} \cdot {\bf \nabla}_p \delta f_k + 
{\bf \nabla}_p \cdot {\left( {\bf F}_R^{(k)} \delta f_k 
\right)} = 
\nonumber 
\end{eqnarray}

\begin{eqnarray}
= - {\left( \delta {\bf F}_B^{(k)} + 
{\bf F}_{N0}^{(k)} \right)} \cdot {\bf \nabla}_p f_{k0}. 
\label{Deltafk} 
\end{eqnarray}

\noindent
The next step consists in defining the weight function that is relative
to the total distribution as

\begin{eqnarray}
W_k {\left( {\bf x}, {\bf p}; s \right)} = 
{\frac {\delta f_k {\left( {\bf x}, {\bf p}; s \right)}} 
{f_k {\left( {\bf x}, {\bf p}; s \right)}}}. 
\label{Weight} 
\end{eqnarray}

\noindent
Substituting 

\begin{eqnarray}
\delta f_k = W_k f_k  \qquad \qquad ; \qquad \qquad 
f_k = {\frac {f_{k0}} {1 - W_k}} 
\label{Relw} 
\end{eqnarray}

\noindent
into (\ref{Deltafk}) and taking into account (\ref{Liouvk}) 
we finally arrive at the evolution equation for the weights: 

\begin{eqnarray}
{\frac {\partial W_k} {\partial s}} + {\bf p} 
\cdot {\bf \nabla}_x W_k + {\left( {\bf F}_L^{(k)} + 
{\bf F}_B^{(k)} + {\bf F}_R^{(k)} \right)} 
\cdot {\bf \nabla}_p W_k = 
\nonumber 
\end{eqnarray}

\begin{eqnarray}
= - {\frac {1} {f_k}} {\left( \delta {\bf F}_B^{(k)} + 
{\bf F}_{N0}^{(k)} \right)} \cdot {\bf \nabla}_p f_{k0} = 
\nonumber 
\end{eqnarray}

\begin{eqnarray}
= {\frac {W_k - 1} {f_{k0}}} {\left( \delta {\bf F}_B^{(k)} + 
{\bf F}_{N0}^{(k)} \right)} \cdot {\bf \nabla}_p f_{k0}. 
\label{Weightk} 
\end{eqnarray}

\noindent
Equation (\ref{Weightk}) can be solved formally by the method of 
characteristics. The first couple of equations for the 
characteristics are precisely the equations of motion (\ref{Lan1}) 
and (\ref{Lan2}). Suppose their solution (particle's trajectory in 
phase space) ${\left\{ {\bf x} (s) \; , \; {\bf p} (s) \right\}}$ 
is known, and let us write down the last one of the equations for 
the characteristics 

\begin{eqnarray}
{\frac {1} {W_k - 1}} {\frac {d W_k} {d s}} = 
{\left. 
{\frac {1} {f_{k0}}} {\left( \delta {\bf F}_B^{(k)} + 
{\bf F}_{N0}^{(k)} \right)} \cdot {\bf \nabla}_p f_{k0} 
\right|}_{{\bf x}, {\bf p} \longrightarrow {\rm trajectory}}. 
\label{Charact} 
\end{eqnarray}

\noindent
Note that its right-hand-side is a function of $s$ only, provided 
${\bf x}$ and ${\bf p}$ are replaced by particle's 
trajectory in phase space ${\left\{ {\bf x} (s) \; , \; 
{\bf p} (s) \right\}}$. Therefore equation (\ref{Charact}) can be 
integrated readily to give:

\begin{eqnarray}
W_k (s) = 1 + {\left[ W_k {\left( s_0 \right)} - 
1 \right]} \exp {\left\{ \int \limits_{s_0}^s 
{\left. {\frac {d \sigma} {f_{k0} (\sigma)}} 
{\left[ \delta {\bf F}_B^{(k)} (\sigma) + 
{\bf F}_{N0}^{(k)} (\sigma) \right]} \cdot 
{\bf \nabla}_p f_{k0} (\sigma) \right|}_{{\bf x} 
(\sigma) \; , \; {\bf p} (\sigma) } \right\}}. 
\label{Solweight} 
\end{eqnarray}

\setcounter{equation}{0}

\section{The Fokker-Planck Equation}

To derive the desired equation let us define the distribution 
function ${\cal F}_{k0} {\left( {\bf x}, {\bf p}; s \right)}$ and 
the fluctuation $\delta f_{k0} {\left( {\bf x}, {\bf p}; s \right)}$ 
according to the relations:

\begin{eqnarray}
{\cal F}_{k0} {\left( {\bf x}, {\bf p}; s \right)} = 
{\left \langle 
f_{k0} {\left( {\bf x}, {\bf p}; s \right)} 
\right \rangle} \qquad ; \qquad 
\delta f_{k0} {\left( {\bf x}, {\bf p}; s \right)} = 
f_{k0} {\left( {\bf x}, {\bf p}; s \right)} - 
{\cal F}_{k0} {\left( {\bf x}, {\bf p}; s \right)}, 
\label{Fluct} 
\end{eqnarray}

\noindent
where $\left \langle \cdots \right \rangle$ implies statistical 
average. Neglecting second order terms and correlators in 
$\delta f_{k0}$ and $\delta f_{(3-k)0}$ that generally give rise to 
collision integrals, we write down the equations for 
${\cal F}_{k0}$ and $\delta f_{k0}$

\begin{eqnarray}
{\frac {\partial {\cal F}_{k0}} {\partial s}} + 
{\bf p} \cdot {\bf \nabla}_x {\cal F}_{k0} + 
{\left( {\bf F}_L^{(k)} + {\bf F}_{L0}^{(k)} \right)} 
\cdot {\bf \nabla}_p {\cal F}_{k0} + 
{\bf \nabla}_p \cdot {\left( {\bar{\bf F}}_R^{(k)} 
{\cal F}_{k0} \right)} = 
\nonumber 
\end{eqnarray}

\begin{eqnarray}
= - {\bf \nabla}_p \cdot 
{\left \langle {\widetilde{\bf F}}_R^{(k)} 
\xi_k (s) \delta f_{k0} \right \rangle}, 
\label{Fcal0} 
\end{eqnarray}

\begin{eqnarray}
{\frac {\partial \delta f_{k0}} {\partial s}} = 
- {\bf \nabla}_p \cdot 
{\left( {\widetilde{\bf F}}_R^{(k)} 
\xi_k (s) {\cal F}_{k0} \right)} + 
O {\left( \delta f_{k0} \right)}, 
\label{Delf0} 
\end{eqnarray}

\noindent
where ${\bar{\bf F}}_R^{(k)}$ and ${\widetilde{\bf F}}_R^{(k)}$ 
denote the deterministic and the stochastic parts of the radiation 
friction force ${\bf F}_R^{(k)}$ respectively. Moreover, the force 
${\bf F}_{L0}^{(k)}$ should be calculated now with the distribution 
function ${\cal F}_{k0}$. Equation (\ref{Delf0}) has a trivial 
solution 

\begin{eqnarray}
\delta f_{k0} (s) = - {\bf \nabla}_p \cdot 
\int \limits_0^{\infty} d \sigma 
{\widetilde{\bf F}}_R^{(k)} (s - \sigma) 
\xi_k (s - \sigma) {\cal F}_{k0} (s - \sigma), 
\label{Delf0sol} 
\end{eqnarray}

\noindent
which is substituted into equation (\ref{Fcal0}) yielding the 
Fokker-Planck equation:

\begin{eqnarray}
{\frac {\partial {\cal F}_{k0}} {\partial s}} + 
{\bf p} \cdot {\bf \nabla}_x {\cal F}_{k0} + 
{\left( {\bf F}_L^{(k)} + {\bf F}_{L0}^{(k)} \right)} 
\cdot {\bf \nabla}_p {\cal F}_{k0} + 
{\bf \nabla}_p \cdot {\left( {\bar{\bf F}}_R^{(k)} 
{\cal F}_{k0} \right)} = 
\nonumber 
\end{eqnarray}

\begin{eqnarray}
= {\bf \nabla}_p \cdot 
{\left[ {\widetilde{\bf F}}_R^{(k)} 
{\bf \nabla}_p \cdot {\left( 
{\widetilde{\bf F}}_R^{(k)} {\cal F}_{k0} 
\right)} \right]}. 
\label{Fokker} 
\end{eqnarray}

In order to carry out the $\delta f$ method effectively, it is important to
find an equilibrium solution of $f_0$ (or very slowly varying solution)
so that the evolution of $\delta f$ is separate in time scale from that
of $f_0$. In the following we discuss the equation and the solution of 
the $f_0$ distribution.

For the sake of simplicity, in what follows bellow in this Section, we 
consider one dimension only (say $x$), since the results can be easily 
generalized to the multidimensional case, provided the x-z coupling is 
neglected. Let us write down the Fokker-Planck equation (\ref{Fokker}) 
in the simplified form:

\begin{eqnarray}
{\frac {\partial {\cal F}_{k0}} {\partial s}} + 
p {\frac {\partial {\cal F}_{k0}} {\partial x}} - 
F_k (s) x 
{\frac {\partial {\cal F}_{k0}} {\partial p}} = 
\Gamma_k {\frac {\partial} {\partial p}}
{\left( p {\cal F}_{k0} \right)} + 
{\cal D}_k 
{\frac {\partial^2 {\cal F}_{k0}} {\partial p^2}}, 
\label{Fokker1} 
\end{eqnarray}

\noindent
where

\begin{eqnarray}
\Gamma_k = {\frac { p_{k0} {\cal C}_1} {2 \pi R}} 
\int \limits_0^{2 \pi R} ds 
{\left| {\bf B}_k (s) \right|}^2 
\qquad ; \qquad 
{\cal D}_k = {\frac {p_{k0}^2 {\cal C}_2} {4 \pi R}} 
\int \limits_0^{2 \pi R} ds 
{\left| {\bf B}_k (s) \right|}^3 
{\left \langle p_{kx}^2 (s) \right \rangle}, 
\label{Damping} 
\end{eqnarray}

\begin{eqnarray}
F_k (s) = {\frac {G_x^{(k)} (s)} {R^2}} + 
\lambda_k \delta_p (s) A_x^{(k)} (s) 
\qquad ; \qquad 
A_x^{(k)} (s) x = {\left. 
{\frac {\partial V_k} {\partial x}} 
\right|}_{{\rm linear \; part}}. 
\label{Coeffic} 
\end{eqnarray}

\noindent
Let us seek for a solution of the Fokker-Planck equation 
(\ref{Fokker1}) in the general form:

\begin{eqnarray}
{\cal F}_{k0} {\left( x, p; s \right)} = 
a_k (s) \exp {\left[ - {\frac 
{{\widetilde{\gamma}}_k (s) x^2 + 
2 {\widetilde{\alpha}}_k (s) x p + 
{\widetilde{\beta}}_k (s) p^2 } 
{2 \epsilon_{x0}^{(k)}}} \right]}, 
\label{Foksol} 
\end{eqnarray}

\noindent
where $\epsilon_{x0}^{(k)}$ is a scaling factor with dimensionality 
and meaning of emittance. Direct substitution of (\ref{Foksol}) 
into (\ref{Fokker1}) and equating similar powers (up to second 
order) in $x$ and $p$ yield the following equations for 
the unknown coefficients:

\begin{eqnarray}
{\frac {d a_k} {d s}} = \Gamma_k a_k 
{\left( 1 - {\frac {{\widetilde{\beta}}_k} 
{\beta_k^{(eq)}}} \right)}, 
\label{Aequk} 
\end{eqnarray}

\begin{eqnarray}
{\frac {d {\widetilde{\alpha}}_k} {d s}} = 
F_k {\widetilde{\beta}}_k - {\widetilde{\gamma}}_k + 
\Gamma_k {\widetilde{\alpha}}_k 
{\left( 1 - {\frac {2 {\widetilde{\beta}}_k} 
{\beta_k^{(eq)}}} \right)}, 
\label{Alequk} 
\end{eqnarray}

\begin{eqnarray}
{\frac {d {\widetilde{\beta}}_k} {d s}} = 
- 2 {\widetilde{\alpha}}_k + 
2 \Gamma_k {\widetilde{\beta}}_k 
{\left( 1 - {\frac {{\widetilde{\beta}}_k} 
{\beta_k^{(eq)}}} \right)}, 
\label{Beequk} 
\end{eqnarray}

\begin{eqnarray}
{\frac {d {\widetilde{\gamma}}_k} {d s}} = 
2 F_k {\widetilde{\alpha}}_k - 
2 \Gamma_k {\frac {{\widetilde{\alpha}}_k^2} 
{\beta_k^{(eq)}}}, 
\label{Gaequk} 
\end{eqnarray}

\noindent
where

\begin{eqnarray}
\beta_k^{(eq)} = {\frac 
{\Gamma_k \epsilon_{x0}^{(k)}} {{\cal D}_k}} 
\label{Betaeq} 
\end{eqnarray}

\noindent
is the {\it equilibrium} $\beta$-function. 

It is important to note that when the damping vanishes 
${\left( \Gamma_k = 0 \right)}$ the above equations are 
exactly the same as the well-known differential equations 
for the Courant-Snyder parameters. In this sense the functions 
${\widetilde{\alpha}}_k$, ${\widetilde{\beta}}_k$ and 
${\widetilde{\gamma}}_k$ can be regarded as a 
generalization of the Courant-Snyder parameters in the case when 
radiation damping and quantum excitation are present. The 
well-known quantity 

\begin{eqnarray}
{\widetilde{\cal I}}_k = \det 
{\left( \begin{array}{cc} 
{\widetilde{\gamma}}_k & 
{\widetilde{\alpha}}_k \\ 
{\widetilde{\alpha}}_k & 
{\widetilde{\beta}}_k 
\end{array} 
\right)} = {\widetilde{\beta}}_k 
{\widetilde{\gamma}}_k - 
{\widetilde{\alpha}}_k^2 
\label{Invar} 
\end{eqnarray}

\noindent
is no longer invariant. It is easy to check that its dynamics 
is governed by the equation

\begin{eqnarray}
{\frac {d {\widetilde{\cal I}}_k} {d s}} = 
2 \Gamma_k {\widetilde{\cal I}}_k 
{\left( 1 - {\frac {{\widetilde{\beta}}_k} 
{\beta_k^{(eq)}}} \right)}. 
\label{Invarequ} 
\end{eqnarray}

\noindent
Comparison between equations (\ref{Aequk}) and 
(\ref{Invarequ}) shows that 

\begin{eqnarray}
a_k (s) = C_{k0} 
{\sqrt{{\widetilde{\cal I}}_k (s)}} 
\label{Compare} 
\end{eqnarray}

\noindent
with $C_{k0}$ an arbitrary constant as it should be. 
Therefore the solution (\ref{Foksol}) takes its final form

\begin{eqnarray}
{\cal F}_{k0} {\left( x, p; s \right)} = 
{\frac {{\sqrt{{\widetilde{\cal I}}_k (s)}}} 
{2 \pi \epsilon_{x0}^{(k)}}} 
\exp {\left[ - {\frac 
{{\widetilde{\gamma}}_k (s) x^2 + 
2 {\widetilde{\alpha}}_k (s) x p + 
{\widetilde{\beta}}_k (s) p^2 } 
{2 \epsilon_{x0}^{(k)}}} \right]}, 
\label{Foksolfin} 
\end{eqnarray}

\noindent
Let us define now the dimensionless {\it envelope} function 
$\sigma_k$ according to the relations 

\begin{eqnarray}
\sigma_k = {\frac {\sqrt{\beta_{ke}}} {a_k}} 
\qquad \qquad ; \qquad \qquad 
\beta_{ke} = {\frac {{\widetilde{\beta}}_k} 
{\beta_k^{(eq)}}}. 
\label{Envelope} 
\end{eqnarray}

\noindent
Manipulating equations (\ref{Alequk}), (\ref{Beequk}) and 
(\ref{Gaequk}) for the generalized Courant-Snyder parameters one can 
eliminate ${\widetilde{\alpha}}_k$ and ${\widetilde{\gamma}}_k$ 
and obtain a single equation for the envelope $\sigma_k$, which 
combined with equation (\ref{Aequk}) comprises a complete set: 

\begin{eqnarray}
{\frac {d^2 \sigma_k} {d s^2}} + \Gamma_k 
{\frac {d \sigma_k} {d s}} + F_k \sigma_k = 
{\frac {1} {\beta_k^{(eq)2} a_k^2 \sigma_k^3}}, 
\label{Envequ} 
\end{eqnarray}

\begin{eqnarray}
{\frac {d a_k} {d s}} = \Gamma_k a_k 
{\left( 1 - a_k^2 \sigma_k^2 \right)}. 
\label{Aequk1} 
\end{eqnarray}

\noindent
By solving equations (\ref{Envequ}) and (\ref{Aequk1}) one can 
obtain a complete information about the evolution of the 
${\cal F}_{k0}$ part of the distribution function. However, solving 
the above system of equations for the beam envelopes and amplitudes 
of the distributions is not an easy task. For that purpose we 
develop in the next Section a numerical scheme which is more suited 
for direct code implementation. 

\section{Numerical Algorithm}

\setcounter{equation}{0}

In the previous Sections, we have established the theoretical 
foundation of the nonlinear $\delta f$ method for the beam-beam 
interaction. In this Section we will apply those results to outline
numerical algorithms suitable for computer simulation.   

Starting with Eq. (\ref{Fequat0}), because the forces in the 
equation both from lattice and the on-coming beam are linear, its 
solution is well known Gaussian distribution (for example as shown 
in the previous Section in the one-dimensional case) 

\begin{equation}
{\cal F}_{k0} ({\bf z}; s) = {\frac {1} {{\left[ 2 \pi 
\det {\left( {\widehat{\Sigma}}_k \right)} 
\right]}^{\frac {3} {2}}}} \exp {\left( -{\frac {1} {2}} 
{\bf z}^T \cdot {\widehat{\Sigma}}_k^{-1} \cdot {\bf z} 
\right)},
\end{equation}

\noindent
where ${\widehat{\Sigma}}_k$ is the matrix of the second moments for 
the distribution and ${\bf z}$ is a vector in the six-dimensional 
phase space. Based on the method of the beam-envelope \cite{ohmi}, 
the propagation of ${\cal F}_{k0}$ can be represented as the 
iteration of the ${\widehat{\Sigma}}_k$ matrix, 

\begin{equation}
{\widehat{\Sigma}}_k^{(i+1)} = {\widehat{\cal M}}_k 
\cdot {\widehat{\Sigma}}_k^{(i)} 
\cdot {\widehat{\cal M}}_k^T + {\widehat{D}}_k,
\end{equation}

\noindent
where ${\widehat{\cal M}}_k$ is the one-turn matrix including the 
linear beam-beam force of the on-coming beam, and the radiation 
damping and ${\widehat{D}}_k$ is the one-turn quantum diffusion matrix. 
Both ${\widehat{\cal M}}_k$ and ${\widehat{D}}_k$ can be extracted 
from the lattice using for example the LEGO code \cite{lego}, 
\cite{quantum}. However, there is a difference  compared to the 
situation of a single storage ring, namely, we have to simultaneously 
iterate the Gaussian distribution for both beams, since the linear 
map ${\widehat{\cal M}}_k$ depends on the beam size of the other beam. 

Combining Eqs. (\ref{Mpsd}) and (\ref{Weight}), the perturbative 
part of the beam distribution $\delta f_k$ has a representation in 
terms of macro-particles 

\begin{eqnarray}
\delta f_k {\left( {\bf x}, {\bf p}; s \right)} = 
{\frac {1} {N_k}} \sum \limits_{n=1}^{N_k} 
W_k^{(n)} (s) \delta 
{\left[ {\bf x} - {\bf x}_n^{(k)} (s) \right]} 
\delta {\left[ {\bf p} - {\bf p}_n^{(k)} (s) 
\right]},
\label{Mpsdd} 
\end{eqnarray}

\noindent
where $W_k^{(n)} (s)$ is the dynamical weight of the $n$-th particle 
from the $k$-th beam. 

As a part of the solution for Eq. (\ref{Weightk}), the propagation of 
the particle coordinates in phase space is the same as the conventional 
PIC code \cite{cai} provided that the beam-beam force is the sum of the
two parts from both ${\cal F}_{k0}$ and $\delta f_k$. 

For the ${\cal F}_{k0}$ part, we can apply the well known 
Erskine-Bassetti formula \cite{bassetti} for a Gaussian beam. The 
force due to the $\delta f_k$ is obtained by solving the two-dimensional 
Poisson equation. In addition to the change of the coordinate, the 
weight of the particle should be propagated according to Eq. 
(\ref{Solweight}). The weight should be updated after the change of 
the coordinate since the change of the weight depends on the trajectory 
of the particle.

\section{ Summary}

We have developed an efficacious algorithm for simulating the beam-beam
interaction in a synchrotron collider with (or without) synchrotron 
radiation. The nonlinear $\delta f$ method has been introduced into
the evolutionary description of subtle changes of the counter streaming
distribution of the colliding beams over many revolutions. The overall
equation that describes this evolution is the Fokker-Planck equation
(with the radiative process and quantum fluctuations). In order to
isolate the $\delta f$ distribution from the average distribution,
we analyze the solution of the Fokker-Planck equation. Obtained is
a form of solution in which the time dependence is parameterized through
a slow evolution (slow compared with the changes in the $\delta f$ distribution
due to the individual beam-beam interaction) in the Courant-Snyder parameters
and the emittance of the beam. This algorithm will enhance the analysis
capability to scrutinize greater details and subtle effects in the 
beam-beam interaction than the PIC version which has been
widely deployed \cite{cai}.

The current algorithm as well as the previous one \cite{cai} have been
developed with an immediate application to the PEP-II B-factory collider.
The code \cite{cai} has already been applied to describe the beam-beam 
interaction in the PEP-II with unprecedented accuracy and reproduction 
faithfulness, and will be sufficient to study the overall dynamics such
as the analysis of resonance instabilities and associated luminosity
functions. It is anticipated, however,  that the numerical noise associated 
with the PIC will require either an inordinate amount of macro-particle
deployment or a level of noise high enough to mask some minute phase space
structure that may manifest in subtle but important long-time evolution
of the beam such as particle diffusion. It is here that the current 
algorithm will cope with the problem. 

\subsection*{Acknowledgments}

We would like to thank John Irwin and Ron Ruth for their continuous 
support and encouragement. It is our pleasure to thank Sam Heifets and
Robert Warnock for many stimulating discussions. One of the
authors (T.T.) is supported in part by DOE contract W-7405-Eng.48 and
DOE grant DE-FG03-96ER40954.

 


\begin{thebibliography}{9}

\bibitem{chao} A.W. Chao, {\it Physics of Collective Beam 
Instabilities in High Energy Accelerators}, Wiley, New York, 
1993.

\bibitem{neuffer} D. Neuffer, A. Riddiford and A. Ruggiero, 
IEEE Trans. Nucl. Sci., {\bf NS-30}, 2430 (1983).

\bibitem{month} M. Month and J.C. Herrera eds., {\it Nonlinear 
Dynamics and the Beam-Beam Interaction}, AIP, New York, 1979.

\bibitem{birdsall} C.K. Birdsall and A.B. Langdon, {\it Plasma 
Physics via Computer Simulation}, McGraw--Hill, New York, 1983.

\bibitem{tajima} T. Tajima, {\it Computational Plasma Physics}, 
Addison--Wesley, Reading, Mass., 1989.

\bibitem{krish1} S. Krishnagopal and R. Siemann, ``Coherent 
Beam-Beam Interaction in Electron-Positron Colliders'', Phys. 
Rev. Lett., {\bf 67}, 2461 (1991).

\bibitem{krish2} S. Krishnagopal, ``Luminosity-Limiting Coherent 
Phenomena in Electron-Positron Colliders'', Phys. Rev. Lett., 
{\bf 76}, 235 (1996).

\bibitem{cai} Y. Cai, A.W. Chao, S.I. Tzenov and T. Tajima, 
``Simulation of the Beam-Beam Effects in $e^+e^-$ Storage Rings 
with a Method of Reducing the Region of Mesh'', SLAC-PUB-8589, 
August 2000.

\bibitem{book} ``PEP-II: An Asymmetric B Factory'', Conceptual Design 
Report, SLAC-418, June 1993.

\bibitem{perkins} T. Tajima and F.W. Perkins, in Proc. of 1983 
Sherwood Theory Meeting, Univ. of Maryland, Arlington, VA, 1983.

\bibitem{kots} M. Kotschenreuther, Bull. Am. Phys. Soc., {\bf 33}, 
2109 (1988).

\bibitem{koga} J.K. Koga and T. Tajima, J. Comput. Phys., 
{\bf 116}, 314 (1995).

\bibitem{klimon} Yu.L. Klimontovich, {\it The Statistical Theory 
of Non-equilibrium Processes in a Plasma}, MIT Press, Cambridge, 
MA, 1967.

\bibitem{ohmi} K. Ohmi, K. Hirata, and K. Oide, ``From the Beam-Envelope 
Matrix to Synchrotron-Radiation Integrals,'' Phys. Rev. E {\bf 49} 751 (1994).

\bibitem{lego} Y. Cai, M. Donald, J. Irwin and Y. Yan, ``LEGO: A
Modular Accelerator Design Code,'' SLAC-PUB-7642, August 1997.

\bibitem{quantum} Y. Cai, ``Simulation of Synchrotron Radiation in an Electron
Storage Ring,'' Proceeding of Advanced ICFA Beam Dynamics Workshop on Quantum
Aspects of Beam Physics, Edited by Pisin Chen (1998).

\bibitem{bassetti} M. Bassetti and G. Erskine, CERN ISR TH/80-06 (1980).

\end{thebibliography}
\end{document}